\documentclass[aps,twocolumn,superscriptaddress,groupedaddress]{revtex4-1}  
\usepackage{graphicx,amssymb}
\usepackage{color,ulem}
\usepackage{bm}  
\usepackage{xcolor}

\newcommand{\eqref}[1]{(\ref{#1})}

\begin{document}
\title{Non-invasive thermometer based on proximity superconductor for ultra-sensitive calorimetry}

\author{Bayan Karimi}
\affiliation{QTF Centre of Excellence, Department of Applied Physics, Aalto University School of Science, P.O. Box 13500, 00076 Aalto, Finland}
\author{Jukka P. Pekola}
\affiliation{QTF Centre of Excellence, Department of Applied Physics, Aalto University School of Science, P.O. Box 13500, 00076 Aalto, Finland}

\date{\today}

\begin{abstract}
We present radio-frequency thermometry based on a tunnel junction between a superconductor and proximitized normal metal. It allows operation in a wide range of biasing conditions. We demonstrate that the standard finite-bias quasiparticle tunneling thermometer suffers from large dissipation and loss of sensitivity at low temperatures, whereas thermometry based on zero bias anomaly  avoids both these problems. For these reasons the latter method is suitable down to lower temperatures, here to about 25 mK. Both thermometers are shown to measure the same local temperature of the electrons in the normal metal in the range of their applicability. 
\end{abstract}

%\pacs{}

\maketitle
%\section{Introduction}

%\begin{figure}[t]
%\centering
%\includegraphics [width=0.9\columnwidth] {Fig1.pdf}
%\caption{Two qubits subjected to (a) correlated and (b) uncorrelated noise sources. 
%\label{fig1}}
%\end{figure}
\section{Introduction}
Thermometry forms a basis of detecting radiation quanta. As such, detection of radiant heat by a thermometer dates back to 1878 by S. P. Langley~\cite{langley,richards}. For measuring energetic quanta, e.g., X-ray photons or radioactive particles, techniques exist for a few decades~\cite{Enss}. As compared, for instance, to observing $6$ keV X-ray photons from Mn K$\alpha$ and K$\beta$ events~\cite{mccammon}, measuring a microwave single photon with about eight orders of magnitude lower energy in the range of $100$ $\mu$eV poses a great challenge~\cite{jukki, nakamura,olesner}. The energy resolution $\delta \epsilon$ of a calorimeter reads $\delta \epsilon=\sqrt{CG_{th}S_T}$, where $C$ denotes the heat capacity of the absorber, $G_{th}$ the thermal conductance to the heat bath, and $S_T$ stands for temperature noise. Among these parameters $S_T$ is directly related to the performance of the thermometer. The challenge is to have a non-invasive thermometer, operating at low enough temperature with noise not exceeding $S_T \sim 10$ $\mu$K$/\sqrt{\rm Hz}$ in order to detect typical $1$ K microwave photons, e.g. in superconducting quantum circuits~\cite{jukki, viisanen}.
\begin{figure}
\centering
\includegraphics [width=\columnwidth] {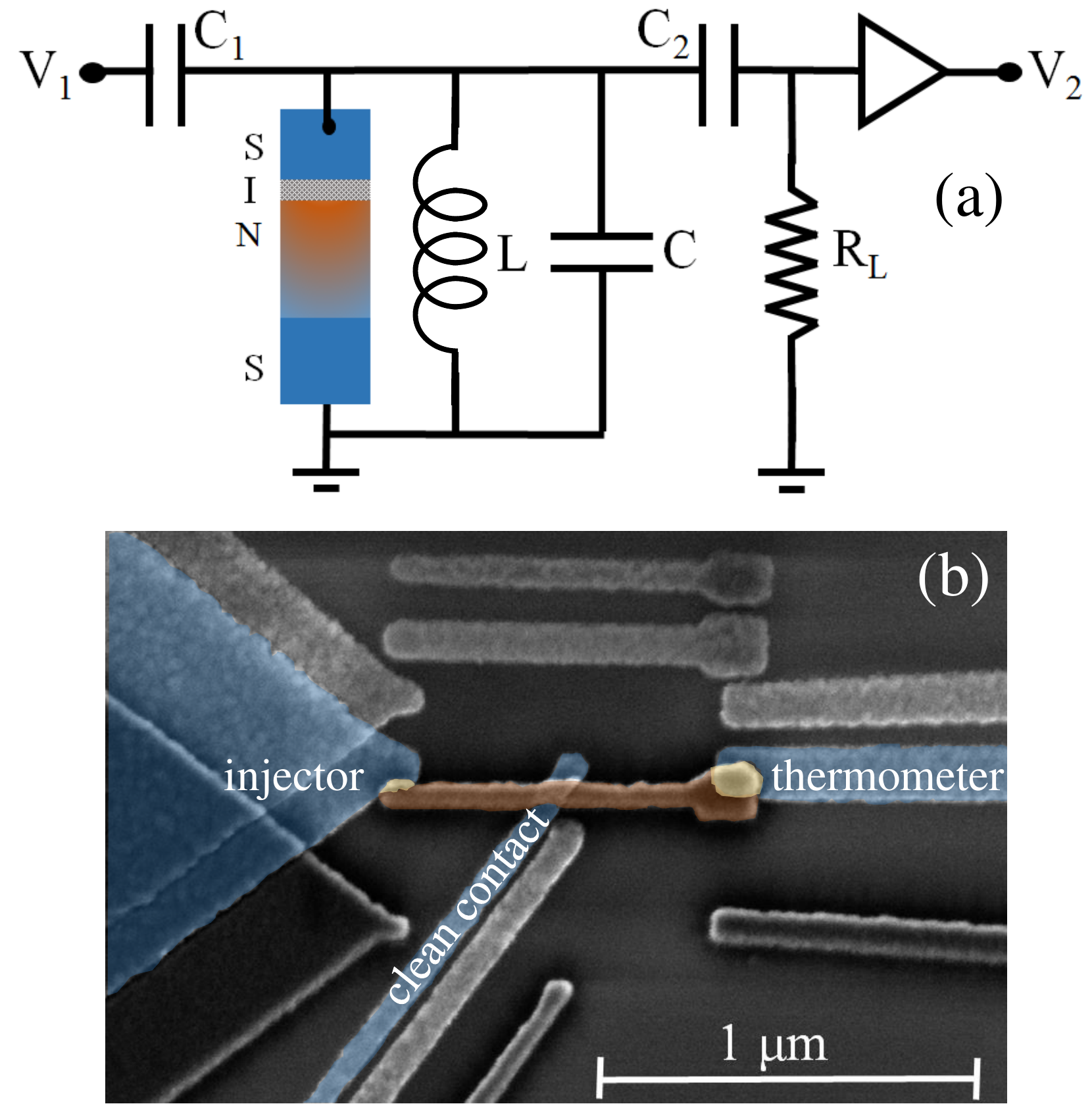}
\caption{(a) The studied setup; schematic illustration of the transmission measurement circuit. $C_1=10.3$ $f$F and $C_2=59.3$ $f$F are the coupling capacitors, $C=0.2$ pF and $L=100$ nH the parameters of the $LC$ resonator, and $R_L=50$ $\Omega$ is the transmission line impedance. (b) Colored scanning electron micrograph of sample A used in this experiment. A normal metallic island (Cu, brown) is in contact with aluminum leads (blue) either via a clean contact or tunnel barrier (light yellow).
\label{fig1}}
\end{figure}
Thermometer candidates for nanocalorimetry purposes include basic NIS tunnel junction probes~\cite{schmidt, Francesco2,simone}, SNS Josephson junctions~\cite{joonas,libin}, SQUIDs (superconducting quantum interference devices)~\cite{Halbertal}, current noise in quantum point contacts~\cite{iftikhar,Banerjee}, Dayem bridges~\cite{foltyn,Francesco}, and proximity circuit QED (Quantum Electro-Dynamics) probes~\cite{olli}. Here, N stands for normal metal, I for insulator barrier, and S for superconductor. The virtue of a NIS junction in a calorimeter is based on its operation in a continuous manner unlike a common switching detector such as a Josephson junction. In this paper we present a thermometer in an RF set-up with about 10 MHz bandwidth; it is a $\mathfrak{N}$IS tunnel junction, where by $\mathfrak{N}$ we denote normal metal influenced by proximity superconductivity~\cite{herve, sophie}. The induced gap in $\mathfrak{N}$ depends exponentially on the distance from the superconductor in the normal metal. In order to make it non-invasive, we monitor the Zero Bias Anomaly (ZBA)~\cite{sophie,kastalsky,peter,alberto} of the junction. We present superior performance as compared to common QuasiParticle (QP) tunneling thermometer due to low dissipation and non-vanishing responsivity down to the lowest temperatures.
\section{Description of the system}\label{sec2}

The studied set-up in this work consists of a small proximitized Cu island coupled to clean superconducting Al contact and to two tunnel junctions. A schematic illustration of the measurement setup and a Scanning Electron Microscope (SEM) image of Sample A are shown in Fig.~\ref{fig1}a and b, respectively. The contact to the right in Fig.~\ref{fig1}b is the thermometer junction, the one to the left is an auxiliary tunnel contact (injector), and the one in the middle is a clean superconductor contact. The structure is fabricated on top of an oxidized silicon substrate by Electron-Beam Lithography (EBL) combined with three-angle shadow evaporation. We present data on two samples. In Sample A (B) the resistance of the thermometer tunnel junction is $R_{\rm T}=8$ k$\Omega$ ($20$ k$\Omega$) and the clean contact is $d=500$ nm ($1$ $\mu$m) away from it. The junction area in both samples is $A\simeq 0.010$ $\mu$m$^2$, yielding specific resistance $R_{\rm T}A=80$ $\Omega\mu$m$^2$ (200 $\Omega\mu$m$^2$) for Sample A (B). In both samples the thickness of the two Al layers (blue color in Fig. \ref{fig1}b) is $20$ nm for both, and $35$ nm for Cu (brown).

In this paper, we focus on the S$\mathfrak N$IS configuration which is used as a RF thermometer. The S$\mathfrak N$ clean contact acts as a heat mirror and fixes the electric potential of the island;  it is directly grounded at the sample stage. Importantly, this contact induces proximity superconductivity at the thermometer junction. In order to obtain fast temperature readout, the superconductor lead of the junction is embedded in an $LC$ resonator, which is connected to input and output rf-lines via coupling capacitors $C_1$ and $C_2$, schematically shown in Fig. \ref{fig1}a. The dc voltage of the thermometer $V_{th}$ is connected to a bias-tee and a small parallel resistor fixed at the printed circuit board of the sample box using a resistive thermocoax dc-line (not shown in Fig. \ref{fig1})~\cite{viisanen}. The $LC$-resonator is made of Al with the thickness of $100$ nm and is fabricated by EBL and one angle metallization; it is placed on a separate chip.  

Using elementary analysis for the circuit in Fig. \ref{fig1}a, and assuming almost all the signal applied on the left is reflected, the ratio of voltages $V_1$ and $V_2$, $s\equiv 2V_2/V_1$, is given by
\begin{widetext}
\begin{equation} \label{e5}
s(\omega)= -\frac{i2\omega^3R_LC_1C_2L}{\{1-\omega^2[L(C+C_1+C_2)+LC_2R_L/R]\}+i\omega[L/R+R_LC_2-\omega^2L(C_1+C)R_LC_2]},
\end{equation}
\end{widetext}
where $R$ is the inverse of the differential conductance $dI/dV$ of the junction and $R_L=50~\Omega$ is the transmission line impedance. For $R_L/R \ll 1$, the resonance frequency $f_0=\omega_0/2\pi \simeq 640$ MHz is given by $\omega_0^2 \approx 1/[L(C_1+C_2+C)]$. Then at resonance
\begin{equation} \label{e7}
s(\omega_0)= -\frac{2C_1}{C_2} \frac{1}{1+R_0dI/dV},
\end{equation}
where $R_0 =  (\omega_0^2R_LC_2^2)^{-1}$. Thus, for $dI/dV\rightarrow 0$, $s(\omega_0) \rightarrow -2 C_1/C_2$, i.e. it obtains the value given by the ratio of the input and output couplings. 
\begin{figure}
\centering
\includegraphics [width=\columnwidth] {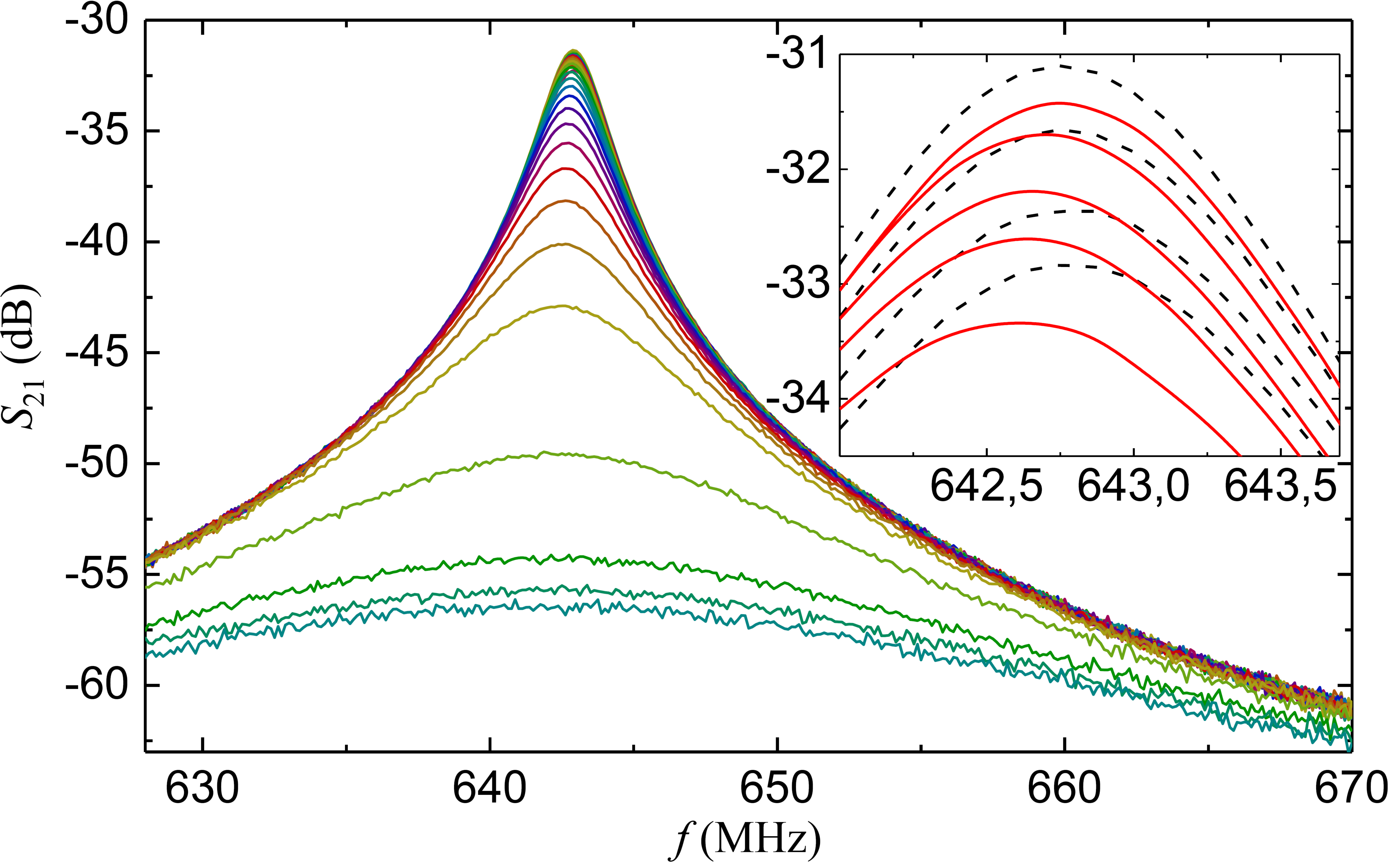}
\caption{The transmission $S_{21}$ measured against the drive frequency $f$ at $T=170$ mK and at $-120$ dBm power. The different curves with the overall trend from top to bottom correspond to thermometer dc bias voltages $V_{th}$ ranging from $0$ to $170$ $\mu$V in $5$ $\mu$V intervals. The inset shows a zoom of a similar measurement at $T=30$ mK where the dashed curves from bottom to top are for $V_{th}=0$, $5$, $10$, $20$ $\mu$V and the solid curves from top to bottom are for $V_{th}=30$, $90$, $105$, $112$, $120$ $\mu$V. We attribute the tiny frequency shift to finite Josephson inductance.
\label{figresonator}}
\end{figure}
Measured in dBm referenced to $P_0=1$ mW, we obtain the transmission in the form 
\begin{eqnarray}\label{S21}
S_{21}(\omega_0)=10 \lg(\frac{V_1^2|s(\omega_0)|^2}{4R_LP_0})= S_0-20 \lg (1+R_0\frac{dI}{dV}),\nonumber\\
\end{eqnarray}
where $S_0=20 \lg (C_1 V_1/(C_2\sqrt{R_LP_0}))$ is a constant offset which in the actual set-up includes also the attenuation and amplification in the lines.
\begin{figure}
\centering
\includegraphics [width=\columnwidth] {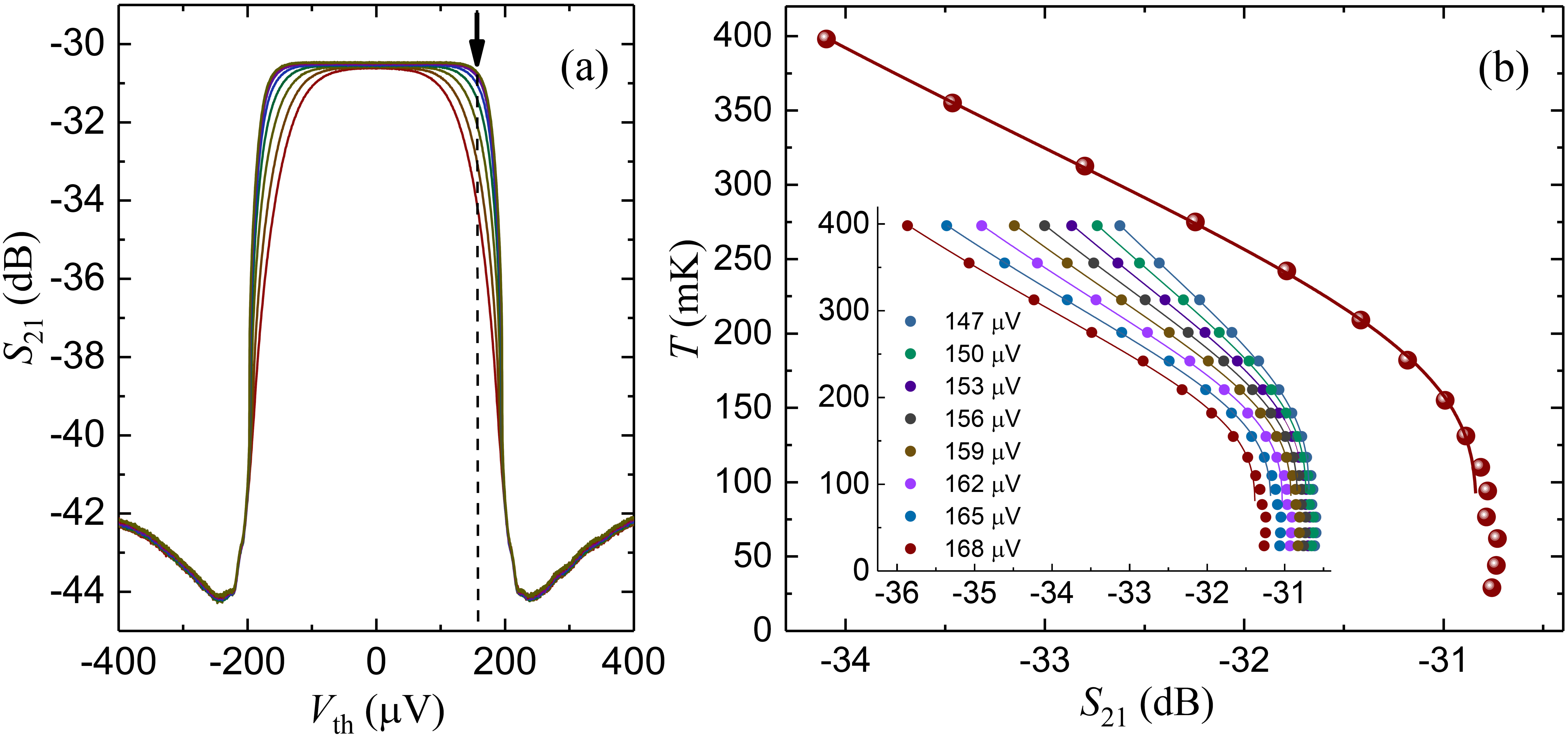}
\caption{Measured characteristics of the RF thermometer in the quasiparticle (QP) regime (Sample B). (a) The transmission $S_{21}$ as a function of the dc bias voltage $V_{th}$. The measurement is performed at temperatures 301, 263, 225, 182, 143, 113, 89, 68, 56, and 50 mK from bottom to top. (b) Temperature calibrations at $V_{th}=156$ $\mu$V measured at -100 dBm of the RF signal in the main panel and at various bias points in the inset. The lines are based on Eq. \eqref{caliT}.
\label{fig2}}
\end{figure}
For low conductance, $dI/dV\ll R_0^{-1}$ we may linearize the relation between $S_{21}$ and $dI/dV$ into the form
\begin{equation}\label{dIdV2}
\frac{dI}{dV}=\frac{\ln (10)}{20R_0}(S_0-S_{21}(\omega_0)),
\end{equation}
i.e. $S_{21}$ measures the negative of the differential conductance of the junction. 

Figure \ref{figresonator} shows the frequency dependence of $S_{21}$ of Sample A, measured between the right and middle contacts of Fig. \ref{fig1}b, around the resonance frequency $f_0$, at the bath temperature $T=170$ mK at different bias voltages $V_{th}$ of the thermometer. (We use the symbol $V_{th}$ interchangably to $V$ when we discuss the actual thermometer junction.) In general, the bias voltage determines the differential conductance $dI/dV$ of the junction in a way to be described below. It is obvious based on the figure that at large biases the resonance line becomes wider due to increasing dissipation (larger $dI/dV$ of the junction). The inset shows a zoom-up of similar data taken at $T=30$ mK, demonstrating a negative frequency shift of about $200$ kHz when biasing the junction away from $V_{th}=0$. This shift is due to the Josephson inductance of the $\mathfrak N$IS junction at zero bias. The Josephson junction with critical current $I_c$ introduces a parallel to $L$ inductance $L_J=\Phi_0/(2\pi I_c)$ leading to a frequency shift $\delta f_0/f_0=L/(2L_J)$ for $L_J \gg L$. Here, $\Phi_0=h/2e$ is the superconducting flux quantum. The measured frequency shift of about $200$ kHz would then imply $I_c \sim 5$ pA for Sample A.

\section{QP thermometry}
Measuring current carried by single electrons (QPs) in a NIS junction has been considered for measurements of power in ultra-sensitive nanobolometers~\cite{nahum,kuzmin}. For an ideal low transparency junction biased at voltage $V$, the expression for QP current reads
\begin{equation}
I=\frac{1}{2eR_T} \int_{-\infty}^\infty dE\,\, n_S(E)\,\{f_N(E-eV)-f_N(E+eV)\},
\end{equation}
where $R_T$ is the resistance of the tunnel junction, $n_S(E)=|E|/\sqrt{E^2-\Delta^2}$ the normalized superconducting density of states for $|E|> \Delta$ and $n_S(E)=0$ for $|E|> \Delta$, and $f_N(E)=(1+\exp(\beta E))^{-1}$ the Fermi distribution in the normal metal at temperature $T=(k_{\rm B}\beta)^{-1}$. Here $\Delta$ denotes the superconducting gap. Far below the critical temperature,  $T\ll T_c$, and for low biases we have 
\begin{equation}\label{II0}
I\approx I_0e^{-(\Delta-eV)/k_BT},
\end{equation}
where $I_0={\sqrt{2\pi k_{\rm B} T\Delta}}/{(2eR_T)}$. Then we can combine Eqs. \eqref{S21} and \eqref{II0} into 
\begin{equation}\label{caliT}
T=\frac{\Delta}{k_B}(1-\frac{eV}{\Delta})\frac{1}{\ln{\tilde r}-\ln ({e^{\frac{\ln 10}{20}(S_0-S_{21})}-1})},
\end{equation}
where ${\tilde r}=\sqrt{{\pi \Delta}/{(2k_{\rm B}T)}}{R_0}/{R_T}$. An estimate of $R_0$ can be obtained from the measured $dI/dV$ versus the bias $V$, because at $|V| \rightarrow \infty$, $dI/dV \rightarrow {R_T}^{-1}$, the resistance of the tunnel junction. On the other hand at low bias $R_0dI/dV \ll 1$. Combining these results we have $R_0/R_T\simeq 10^{{\Delta S}/{20}}-1$, where $\Delta S=S_{21}(V\approx 0)-S_{21}(|V| \gg \Delta /e)$. 

In order to have pure NIS configuration, the direct superconducting contact is placed $1$ $\mu$m away from the thermometer junction in Sample B. In practice, this leads to vanishing proximity effect at the thermometer junction. The transmission $S_{21}$ as a function of voltage bias of the thermometer for a set of bath temperatures $T_{bath}$ is presented in Fig. \ref{fig2}a. In all these measurements bath temperatures are obtained based on primary Coulomb Blockade Thermometer (CBT)~\cite{jukki2}. Figure \ref{fig2}b shows the data points extracted from the transmission curves at fixed voltage biases for different bath temperatures. The solid lines show fits to the corresponding experimental data based on Eq. \eqref{caliT}. It is clear that all the experimental sets match the calculated ones with $\tilde r$ and $S_0$ as fitting parameters. The QP thermometer loses its sensitivity at low $T$ demonstrated by the vanishing responsivity $\mathbb R\equiv |dS_{21}/dT|$. 

It is advantageous for calorimetry to work at as low temperature as possible. This is because the energy resolution of an ideal calorimeter limited by fundamental thermal fluctuations is given by $\delta \epsilon=\sqrt{k_{\rm B}C}T \propto T^{3/2}$~\cite{Landau}. Therefore one would hope to have a sensitive and reliable thermometer down to the lowest temperatures reachable by a standard dilution refrigerator. For this purpose we next present and analyze a different concept, which avoids the vanishing responsivity at low $T$.
\begin{figure}
\centering
\includegraphics [width=\columnwidth] {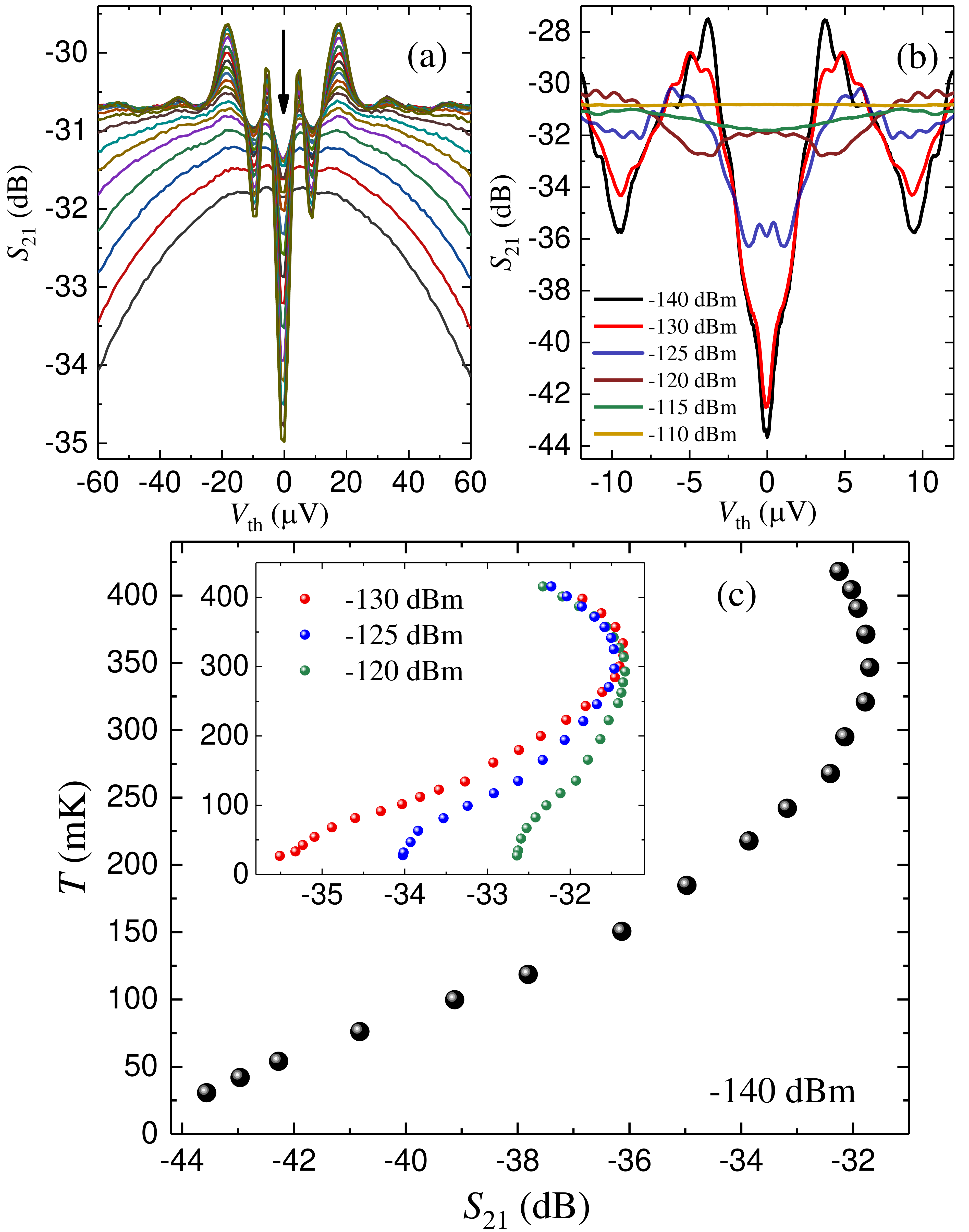}
\caption{The transmission of the proximitized junction of Sample A in the zero bias anomaly (ZBA) regime, $S_{21}$ versus $V_{th}$, for (a) a set of bath temperatures $T_{bath}$ in the range of $27$ to $398$ mK and fixed power $-130$ dBm (b) few power levels at $T=30$ mK. (c) Temperature calibration curves  of the transmitted power at zero bias measured at -140 dBm of the RF signal in the main panel, and at different power levels $-130$, $-125$, and $-120$ dBm in its inset.}
\label{fig3}
\end{figure}
\begin{figure}[t!]
\centering
\includegraphics [width=\columnwidth] {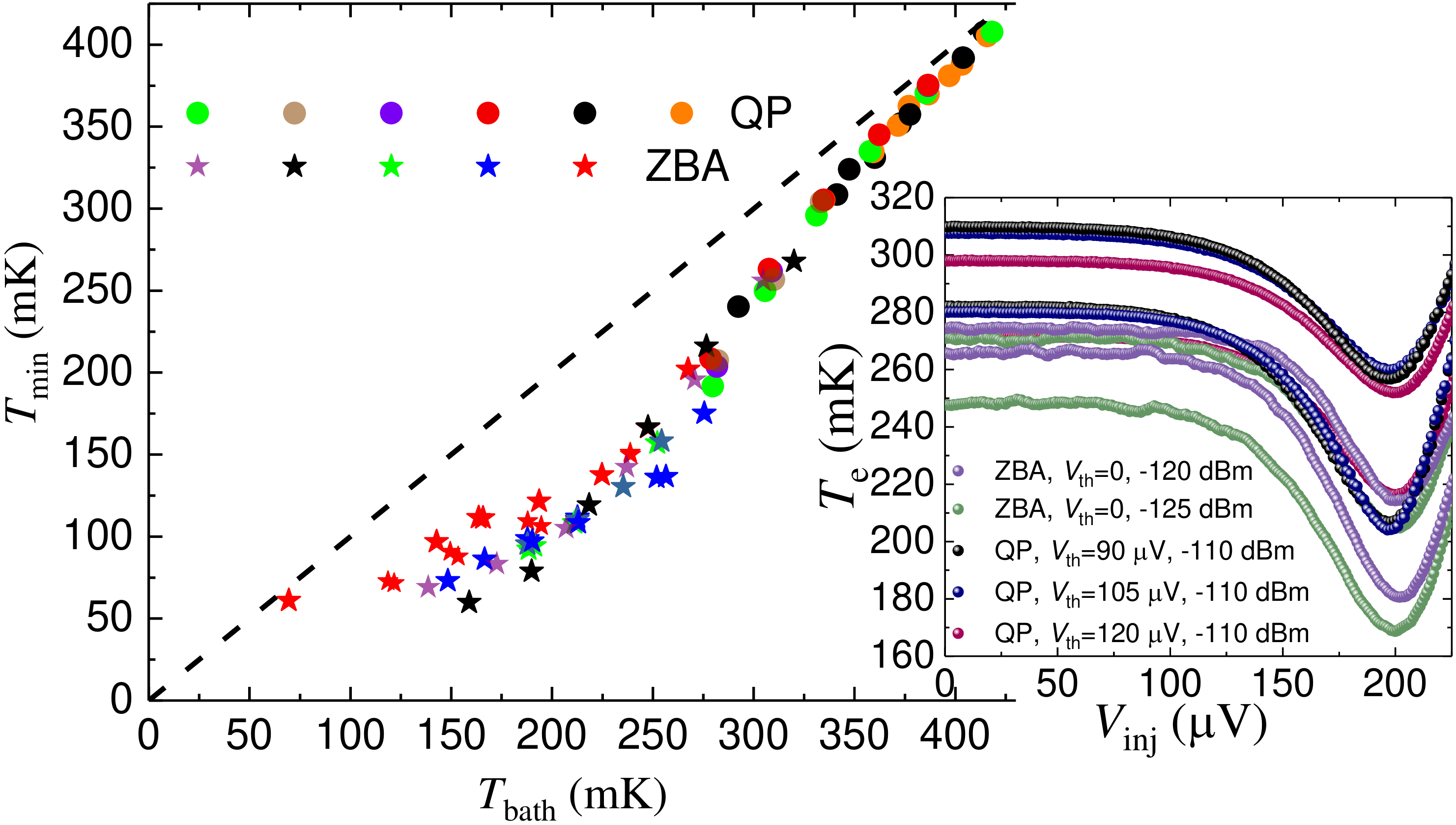}
\caption{Measurement of electron temperature $T_e$ of the absorber under non-equilibrium conditions produced by applying voltage $V_{inj}$ across the injector junction. (right) Voltage dependence of $T_e$ demonstrates cooling, measured with different biasing and power levels of the thermometer. (left) Extracted minimum temperature of $T_e$ as a function of bath temperature $T_{bath}$. Dashed line denotes $T_{min}=T_{bath}$. The symbols in the main figure refer to: $\color{green}{\bullet}$ QP, $V_{th}=90$ $\mu$V, -110 dBm, $\color{brown}{\bullet}$ QP, $V_{th}=90$ $\mu$V, -110 dBm,  $\color{violet}{\bullet}$ QP, $V_{th}=105$ $\mu$V, -110 dBm, $\color{red}{\bullet}$ QP, $V_{th}=120$ $\mu$V, -110 dBm, $\color{black}{\bullet}$ QP, $V_{th}=0$, -120 dBm, $\color{orange}{\bullet}$ QP, $V_{th}=0$, -125 dBm, $\color{violet}{\bigstar}$ ZBA, $V_{th}=20$ $\mu$V, -120 dBm, $\color{black}{\bigstar}$ ZBA, $V_{th}=20$ $\mu$V, -125 dBm, $\color{green}{\bigstar}$ ZBA, $V_{th}=0$, -115 dBm, $\color{blue}{\bigstar}$ ZBA, $V_{th}=0$, -120 dBm, $\color{red}{\bigstar}$ ZBA, $V_{th}=0$, -125 dBm.} 
\label{fig4}
\end{figure}   

\section{Zero Bias Anomaly (ZBA) thermometer} \label{section4}
In Sample A we place the clean superconducting contact closer to the thermometer junction ($d=500$ nm). This gives the proximity effect to the normal metal with superconducting properties extending all the way to the position of the thermometer $\mathfrak{N}$IS junction. In this case a zero bias anomaly (ZBA) arises as shown in Fig. \ref{fig3}a. This is to be compared to the measurement in the same set-up in Fig. \ref{fig2}a, where on Sample B we observe no structure in the low bias region. 

Figure \ref{fig3}b exhibits transmission $S_{21}$ results at different RF power levels  on Sample A. It is vivid that due to the narrow ZBA feature the result is sensitive to the applied power. This ZBA offers us a sensitive probe of temperature here down to the base temperature of the measurement ($25$ mK) as shown in Fig. \ref{fig3}c, where temperature versus the zero-bias $S_{21}$ is presented. This probe is non-dissipative (zero bias) and sensitive (no saturation) at low temperatures, in contrast to the QP thermometer. It is worth pointing out that the responsivity $\mathbb{R}$ of ZBA thermometer at $T<200$ mK is $\simeq 0.06$ dB/mK, which clearly exceeds $\mathbb R \simeq 0.01$ dB/mK of the QP-thermometer in the temperature range of its applicability. Temperature calibrations measured with different powers, shown in the inset of Fig. \ref{fig3}c, demonstrate that measuring with larger power (-120 dBm, -125 dBm) leads to saturation of $S_{21}$ at low temperatures. One can also see that ZBA is composed of several peaks with origin in supercurrent, and possibly in multiple Andreev reflection due to the relatively high transparency of the junction~\cite{tinkham2,Averin,scheer,cuevas}. The back-bending at $T>300$ mK is due to QP current.

In order to compare different thermometers in Sample A, we manipulate the electronic temperature $T_e$ of the N island by applying a bias voltage $V_{inj}$ across the auxiliary junction, denoted "injector" in Fig. \ref{fig1}b with tunnel resistance $R_{inj}\simeq 50$ k$\Omega$. The influence of this bias is the feature depicted in the inset of Fig. \ref{fig4}. In all curves the electronic temperature drops at $V_{inj}\sim \Delta/e$ due to the well-known quasiparticle cooling effect~\cite{jukkireview}. Operating the thermometer at three different conditions including zero with two different powers and several bias values as a QP thermometer indicates close to identical temperatures over the whole $V_{inj}$ bias range. In the main panel of Fig. \ref{fig4}, we have collected measurements with different bias points $V_{th}$ of the thermometers, while measuring the minimum temperatures $T_{min}$ at $V_{inj}\sim \Delta/e$ versus bath temperature $T_{bath}$. QP thermometer at various bias points spans over the range from $\gtrsim$100 to 400 mK, whereas ZBA thermometer can be used down to the minimum temperature of the cryostat in this experiment. 
\begin{figure}
\centering
\includegraphics [width=\columnwidth] {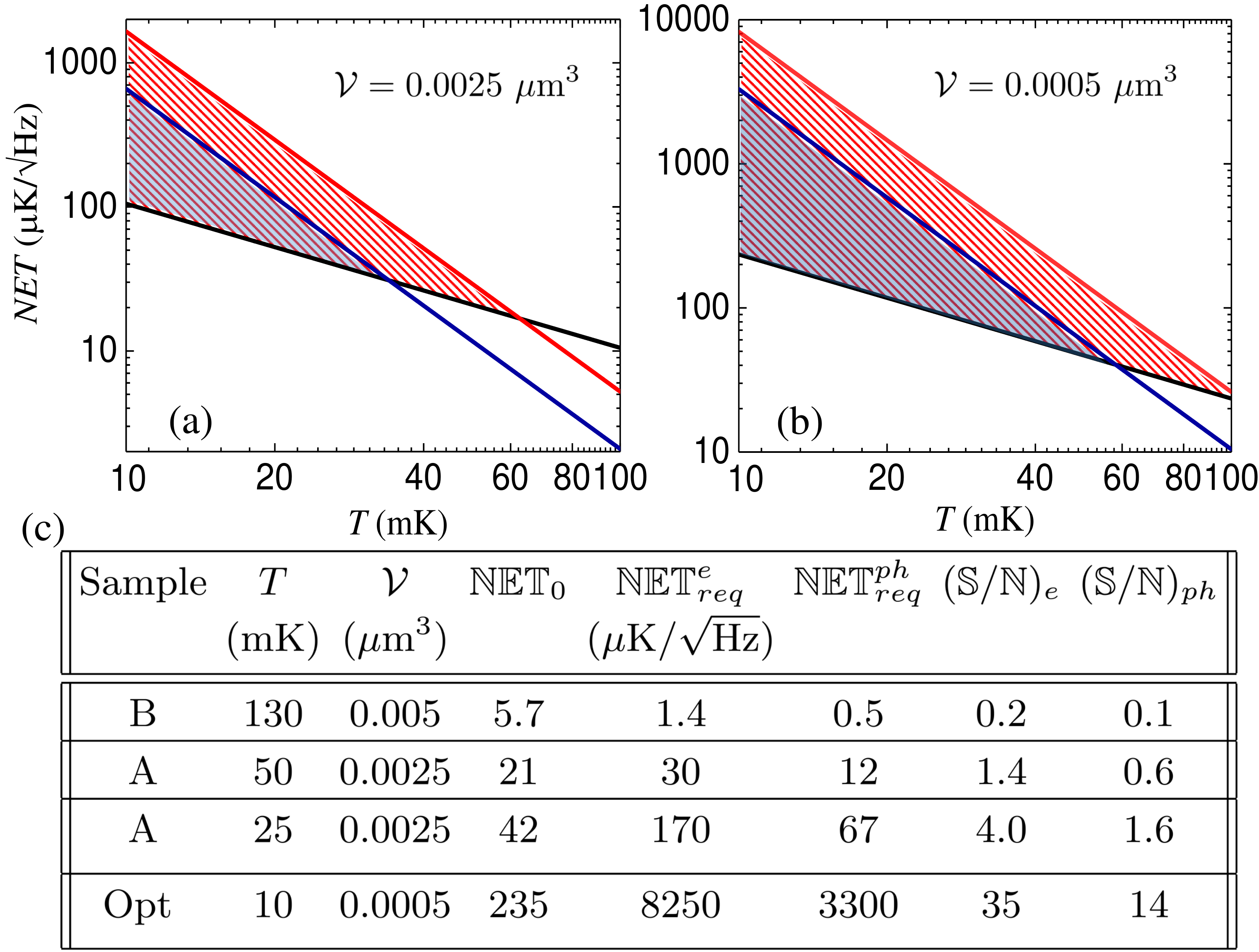}
\caption{Requirements for detecting the heat produced by single electron $\delta \epsilon_e=2.5$ K$\times k_{\rm B}$ and single photon $\delta \epsilon_{ph}=1 $K$\times k_{\rm B}$ quanta. Dependence of various noise-equivalent temperatures $\mathbb{NET}$ on bath temperature $T_{bath}$ for two different volumes of copper absorber, (a) $0.0025~~\mu{\rm m}^3$ and (b) $0.0005~~\mu{\rm m}^3$. In both panels, black line demonstrates the fundamental temperature fluctuations $\mathbb{NET}_0$, red line shows the required noise-equivalent temperature for electrons $\mathbb{NET}_{req}^e$ and blue lines for photons $\mathbb{NET}_{req}^{ph}$. The shaded areas indicate the feasible regimes where the corresponding quanta can be observed. In panel (c), we present concrete examples referring to the Samples A, B and Opt with a small volume yet experimentally feasible. QP and ZBA thermometry are used for samples B and A, respectively, at the minimum temperature $130$ mK for QP and two different temperatures for ZBA. The last two columns give the corresponding signal to noise ratio for detecting electrons $(\mathbb{S}/\mathbb{N})_e$ and photons $(\mathbb{S}/\mathbb{N})_{ph}$. The parameters used for evaluating the present results in panels a-c are: $\Sigma=2\times 10^9 ~{\rm WK^{-5}m^{-3}}$ and $\gamma=70~{\rm JK^{-2}m^{-3}}$. 
\label{fig5}}
\end{figure} 
We observe that the measurements form a continuous set of data showing the optimum cooling around 200 mK. Based on the collapse of the different sets of data in Fig. \ref{fig4} we conclude that all these thermometers measure, in a consistent way, the temperature of the electrons in the island. 

\section{Discussion and conclusions}
Here we discuss feasibility of measuring single energy quanta by ZBA thermometer, and compare its performance to QP method. We show quantitatively that the main advantages of ZBA based calorimetry are its low operation temperature and ultra low power dissipation. The challenge, on the other hand, is that only low input power levels are feasible as the ZBA peak is very narrow in $V_{th}$.

{\it Non-invasiveness of ZBA} - An important key for a nanocalorimeter in order to be able to detect single quanta like photons with energy $\sim 100$ $\mu$V, is to be non-invasive~\cite{fredrik}. Using Eq.~\eqref{II0} and considering $\Delta /e=200$ $\mu$V for superconducting gap of Al, $V=0.8\Delta/e$ and $T=130$ mK as parameters for Sample B, QP thermometer injects $\sim 15$ aW heat. On the other hand, the corresponding power for ZBA thermometer in Sample A (using Eq.~\eqref{dIdV2} and $-140$ dBm for applied RF power) is $\sim 4$ zW at $25$ mK, i.e., $\sim 10^6$ times less than in the QP thermometer.

{\it Sensitivity at low $T$} - We start by estimating the required noise-equivalent temperature $\sqrt{S_T}$ which we denote $\mathbb{NET}_{req}$ in $\mu {\rm K}/\sqrt{{\rm Hz}}$. For energy we choose $\delta \epsilon_e=2.5$ K$\times k_{\rm B}$ for measurements described in~\cite{fredrik} to detect an electron tunneling over the superconducting gap of Al, and $\delta \epsilon_{ph}=1$ K$\times k_{\rm B}$ for a $20$ GHz single microwave photon.
Figures \ref{fig5}a and \ref{fig5}b demonstrate the feasibility of measuring the energy deposited by single electrons and single microwave photons by the envisioned calorimeter with copper absorber. Both the panels present various predicted $\mathbb{NET}$-values as functions of the operation temperature for two different volumes of the absorber, $0.0025$ $\mu {\rm m}^3$ which is the current Sample A, and  $0.0005$ $\mu {\rm m}^3$ that represents a technically realistic tiny absorber (called Sample Opt), respectively. In this figure, the black line in both panels \ref{fig5}a and \ref{fig5}b represents fundamental temperature fluctuations $\mathbb{NET}_0=\sqrt{S_{\dot{Q}}/G_{th}^2}=\sqrt{2k_{\rm B}/(5\Sigma \mathcal{V})}T^{-1}$, where, according to fluctuation dissipation theorem, $S_{\dot{Q}}=2k_{\rm B}T^2G_{th}$ is the heat current noise in equilibrium. The two other lines denote the required noise-equivalent temperature $\mathbb{NET}_{req}^i=\delta \epsilon_i/(\mathcal{V}\sqrt{5\gamma \Sigma})~T^{-5/2}$ for $i=e,ph$, to observe quanta of energy for electrons (red line) and photons (blue line). For our estimations, we use the well-known expressions for thermal conductance to the phonon bath $G_{th}=5\Sigma \mathcal{V} T^4$, and $C=\gamma \mathcal{V} T$ for heat capacity of the normal metal. Here, $\Sigma$ denotes the electron-phonon constant, $\mathcal{V}$ the volume of the island, and $\gamma$ refers to the Sommerfeld constant for electrons in metal. The shaded areas delineate the favourable regimes for detecting these particles. The upper boundary is given by the required $\mathbb{NET}_{req}^i$ and the lower bound (black line) represents the fundamental temperature fluctuations $\mathbb{NET}_0$. 
The table presented in Fig. \ref{fig5}c shows examples of these estimates under four different conditions. The first row presents Sample B with QP thermometry around its lowest operation temperature ($130$ mK). The estimates for ZBA thermometer of Sample A at two different temperatures ($50$ and $25$ mK) are given in the second and third rows, respectively. Here, in the absence of precise temperature calibration for ZBA, the two rows represent the conservative and optimistic estimates of the actual base temperature using ZBA. The fourth row indicates a further optimized sample, Opt, with experimentally feasible target parameters. The fifth and sixth columns in the table, $\mathbb{NET}_{req}^e$ and $\mathbb{NET}_{req}^{ph}$, demonstrate that, it is next to impossible to detect a single electron $e$ or photon $ph$ by QP thermometer. Yet using ZBA at lower $T$, the requirement for $\mathbb{NET}$ is relaxed by one to two orders of magnitude. The typical $\mathbb{NET}$ in the present measurement is around $\sim 30$ $\mu$K/$\sqrt{\rm Hz}$ based on ZBA thermometer (including the amplifier noise). Thus the current $\mathbb{NET}$ is just about to be sufficient to detect $2.5$ K$\times k_{\rm B}$. In the last two columns we present the expected signal to noise ratio $(\mathbb{S}/\mathbb{N})_i$ for different quanta, $i=e,ph$, demonstrating the possibility of detecting them by ZBA at low temperatures and with small absorber.

Finally, we comment briefly on how to optimize the ZBA thermometer in future. The critical current $I_c$ of the $\mathfrak N$IS junction is expected to increase exponentially when the distance of the clean contact $d$ from the junction decreases~\cite{tinkham, Herve2}. Therefore higher responsivity is expected for smaller values of $d$: in practice $d$ can be decreased down to $\sim 50$ nm from the current $500$ nm. Moreover, the frequency shift of the resonance due to Josephson inductance will also be enhanced in this case, giving an extra boost in the sensitivity of the ZBA thermometer. With the low operation temperature ($\sim 10$ mK) and the proposed improvements in design, the ZBA calorimeter can detect single microwave photons.

\section*{acknowledgments}
This work was funded through Academy of Finland grants 297240, 312057 and 303677 and from the European Union's Horizon 2020 research and innovation programme under the European Research Council (ERC) programme and Marie Sklodowska-Curie actions (grant agreements 742559 and 766025). We acknowledge the facilities and technical support of Otaniemi research infrastructure for Micro and Nanotechnologies (OtaNano). We acknowledge J. T. Peltonen and E. T. Mannila for technical support, and F. Brange, P. Samuelsson, D. Golubev, A. Mel'nikov, O-P. Saira, K. L. Viisanen, W. Belzig, I. Khaymovich and P. Muratore-Ginanneschi for useful discussions.


\begin{thebibliography}{99}
%Intro history of thermometry
\bibitem{langley}S. P. Langley, The bolometer, Nature {\bf 57}, 620-622 (1898).

\bibitem{richards}P. L. Richards, Bolometers for infrared and millimeter waves, J. Appl. Phys. {\bf 76}, 1 (1994).

\bibitem{Enss} Christian Enss (Ed), {\it Cryogenic particle detection} (Springer-Verlag Berlin, Heidelberg, 2005).

\bibitem{mccammon} D. McCammon, S. H. Moseley, J. C. Mather, and R. F. Mushotzky, Experimental tests of a single-photon calorimeter for x-ray spectroscopy, J. Appl. Phys. {\bf 56}, 1263 (1984).

\bibitem{jukki} J. P. Pekola, P. Solinas, A. Shnirman, and D. V. Averin, Calorimetric measurement of work in a quantum system, New J. Phys.  {\bf 15}, 115006 (2013).

\bibitem{nakamura} Kunihiro Inomata, Zhirong Lin, Kazuki Koshino, William D. Oliver, Jaw-Shen Tsai, Tsuyoshi Yamamoto, and Yasunobu Nakamura, Single microwave-photon detector using an artificial $\Lambda$-type three-level system, Nature Communications {\bf 7}, 12303 (2016).

\bibitem{olesner} G. Oelsner, C. K. Andersen, M. Reh\'ak, M. Schmelz, S. Anders, M. Grajcar, U. H\"ubner, K. M{\o}lmer, and E. Il{'}ichev, Detection of Weak Microwave Fields with an Underdamped Josephson Junction, Phys. Rev. Appl. {\bf 7}, 014012 (2017).

\bibitem{viisanen} K. L. Viisanen and J. P. Pekola, Anomalous electronic heat capacity of copper nanowires at sub-kelvin temperatures, Phys. Rev. B  {\bf 97}, 115422 (2018).
 
\bibitem{schmidt}D. R. Schmidt, C. S. Yung, and A. N. Cleland, Nanoscale radio-frequency thermometry, Appl. Phys. Lett. {\bf 83}, 1002 (2003).

\bibitem{Francesco2}Francesco Giazotto and Mar\'ia Jos\'e Mart\'inez-P\'erez, The Josephson heat interferometer, Nature {\bf 492}, 401 (2012). 

\bibitem{simone} S. Gasparinetti, K. L. Viisanen, O.-P. Saira,  T. Faivre, M. Arzeo, M. Meschke, and J. P. Pekola, Fast electron thermometry towards ultra-sensitive calorimetric detection. Phys. Rev. Applied {\bf 3}, 014007 (2015).

\bibitem{joonas}J. Govenius, R. E. Lake, K. Y. Tan, and M. M\"ott\"onen, Detection of Zeptojoule Microwave Pulses Using Electrothermal Feedback in Proximity-Induced Josephson Junctions, Phys. Rev. Lett. {\bf 117}, 030802 (2016).

\bibitem{libin} L. B. Wang, O.-P. Saira, and  J. P. Pekola, Fast thermometry with a proximity Josephson junction, Appl. Phys. Lett. {\bf 112}, 013105 (2018). 

\bibitem{Halbertal}Dorri Halbertal, Moshe Ben Shalom, Aviram Uri, Kousik Bagani, Alexander Y. Meltzer, Ido Marcus, Yuri Myasoedov, John Birkbeck, Leonid S. Levitov, Andre K. Geim, Eli Zeldov, Imaging resonant dissipation from individual atomic defects in graphene, Science {\bf 358}, 1303 (2017).

\bibitem{iftikhar} Z. Iftikhar, A. Anthore, S. Jezouin, F.D. Parmentier, Y. Jin, A. Cavanna, A. Ouerghi, U. Gennser, and F. Pierre, Primary thermometry triad at $6$ mK in mesoscopic circuits, Nature Communications {\bf 7}, 12908 (2016).

\bibitem{Banerjee}Mitali Banerjee, Moty Heiblum, Amir Rosenblatt, Yuval Oreg, Dima E. Feldman, Ady Stern and Vladimir Umansky, Observed quantization of anyonic heat flow, Nature {\bf 545}, 22052 (2017). 

\bibitem{foltyn}M. Foltyn and M. Zgirski, Gambling with Superconducting Fluctuations, Phys. Rev. Appl. {\bf 4}, 024002 (2015).

\bibitem{Francesco}Federico Paolucci, Giorgio De Simoni, Elia Strambini,y Paolo Solinas, and Francesco Giazotto, Ultra-efficient superconducting Dayem bridge field-effect transistor, arXiv:1803.04925 (2018). 

\bibitem{olli}O.-P. Saira, M. Zgirski, K.L. Viisanen, D.S. Golubev, and J.P. Pekola, Dispersive thermometry with a Josephson junction coupled to a resonator, Phys. Rev. Appl. {\bf 6}, 024005 (2016).

\bibitem{herve}B. Pannetier and H. Courtois, Andreev reflection and proximity effect, J. Low Temp. Phys. {\bf 118}, 599 (2000).

\bibitem{sophie}S. Gueron, H. Pothier, Norman O. Birge, D. Esteve, and M. H. Devoret, Superconducting Proximity Effect Probed on a Mesoscopic Length Scale. Phys. Rev. Lett. {\bf 77}, 3025 (1996).

\bibitem{kastalsky} A. Kastalsky, A. W. Kleinsasser, L. H. Greene, R. Bhat, F. P. Milliken, and J. P. Harbison, Observation of Pair Currents in Superconductor-Semiconductor Contacts, Phys. Rev. Lett. {\bf 67}, 3026 (1991).

\bibitem{peter}H. A. Nilsson, P. Samuelsson, P. Caroff, and H. Q. Xu, Supercurrent and Multiple Andreev Reflections in an InSb Nanowire Josephson Junction, Nano Lett. {\bf 12}, 228 (2012).

\bibitem{alberto}Alberto Ronzani, Carles Altimiras, and Francesco Giazotto, Highly Sensitive Superconducting Quantum-Interference Proximity Transistor, Phys. Rev. Appl. {\bf 2}, 024005 (2014).

\bibitem{nahum}M. Nahum and John M. Martinis, Ultrasensitive-hot-electron microbolometer, Appl. Phys. Lett. {\bf 63}, 3075 (1993). 

\bibitem{kuzmin}A. V. Gordeeva, V. O. Zbrozhek, A. L. Pankratov, L. S. Revin, V. A. Shamporov, A. A. Gunbina, and L. S. Kuzmin, Observation of photon noise by cold-electron bolometers, Appl. Phys. Lett. {\bf 110}, 162603 (2017). 

\bibitem{jukkireview}Francesco Giazotto, Tero T. Heikkil\"a, Arttu Luukanen, Alexander M. Savin, and Jukka P. Pekola, Opportunities for mesoscopics in thermometry and refrigeration: Physics and applications, Rev. Mod. Phys. {\bf 78}, 217 (2006).

\bibitem{jukki2} J. P. Pekola, K. P. Hirvi, J. P. Kauppinen, and M. A. Paalanen, Thermometry by Arrays of Tunnel Junctions, Phys. Rev. Lett. {\bf 73}, 2903 (1994).

\bibitem{fredrik} F. Brange, P. Samuelsson, B. Karimi, and J. P. Pekola, Nanoscale Quantum Calorimetry with Electronic Temperature Fluctuations, arXiv:1805.02728 (2018).

\bibitem{Landau} E. M. Lifshitz and L. P. Pitaevskii, {\it Statistical Physics, Part 1}, 3rd ed. (Pergamon press, Oxford 1980). 

\bibitem{tinkham2} T. M. Klapwijk, G. E. Blonder, and M. Tinkham, Explanation of subharmonic energy gap structure in superconducting contacts, Physica B \& C {\bf 109}, 1657 (1982). 

\bibitem{Averin} Athanassios Bardas and Dmitri V. Averin, Electron transport in mesoscopic disordered superconductor –normal-metal –superconductor junctions, Phys. Rev. B {\bf 56}, R8518 (1997). 

\bibitem{scheer} E. Scheer, P. Joyez, D. Esteve, C. Urbina, and M. H. Devoret, Conduction channel transmissions of atomic-size aluminum contacts, Phys. Rev. Lett. {\bf 78}, 3535 (1997).

\bibitem{cuevas} J. C. Cuevas, J. Hammer, J. Kopu, J. K. Viljas, and M. Eschrig, Proximity effect and multiple Andreev reflections in diffusive superconductor–normal-metal–superconductor junctions, Phys. Rev. B {\bf 73}, 184505 (2006). 

\bibitem{tinkham} M. Tinkham, {\it Introduction to superconductivity}, 2nd ed. (Dover New York, 1996). 

\bibitem{Herve2} P. Dubos, H. Courtois, B. Pannetier, F. K. Wilhelm, A. D. Zaikin, and G. Sch\"on, Josephson critical current in a long mesoscopic S-N-S junction, Phys. Rev. B {\bf 63}, 064502 (2001). 


\end{thebibliography}
\end{document}